\documentclass{aa}  

\usepackage{graphicx}
\usepackage{txfonts}
\usepackage{color}
\usepackage[]{hyperref}
\begin{document} 

\title{The peculiar ejecta of the nova V1425 Aquilae}

\author{C. Tappert\inst{1}
\and
L. Celed\'on\inst{1}
\and
L. Schmidtobreick\inst{2}
}

\institute{Instituto de F\'isica y Astronom\'ia,
Universidad de Valpara\'iso, Valpara\'iso, Chile\\
\email{claus.tappert@uv.cl}
\and
European Southern Observatory, Santiago, Chile
}

\date{Received XXX; accepted XXX}

\abstract{
Many important details of the mechanisms underlying the ejection of material
during a (classical) nova eruption are still not understood. Here we present
optical spectroscopy and narrow-band images of the nova V1425 Aql, 23 years
after the nova eruption. We find that the ejecta consist of two significantly
different components. The first resembles what is commonly seen in
novae, that is, a symmetric distribution centred on the position of the underlying
cataclysmic binary and presenting both allowed (hydrogen and helium) and
forbidden ([O{\sc iii}] and [N{\sc ii}]) transitions. The second one,
on the other hand, consists of material travelling at an  approximately three times higher velocity that 
is not visible in the allowed transitions, presents a significantly different
[N{\sc ii}] - [O{\sc iii}] ratio, and is located at approximately 2.3 arcsec 
to the southwest of the position of the binary. Comparing the
velocities and spatial extensions of the two ejecta, we find that both
originated in the same nova eruption. We explore possible extrinsic and 
intrinsic
mechanisms for the asymmetry of the high-velocity material in the form
of asymmetrically distributed interstellar material and magnetic accretion,
respectively, but find the available data to be inconclusive. From the 
expansion parallax, we derive a distance for the nova of 3.3(3) kpc.
}


\maketitle
%

\section{Introduction \label{intro_sec}}

In cataclysmic binary stars (CVs), a white dwarf accretes matter from a 
late-type main sequence star. The material is transported either via an 
accretion disc and a boundary layer onto the white dwarf (in the non-magnetic 
dwarf novae and nova-likes) or is funnelled along the magnetic field lines 
directly onto its magnetic pole or poles (in CVs with a strong
magnetic field, so-called polars). An intermediate configuration where the
magnetic field disrupts the inner part of the accretion disc is known as
an intermediate polar. For comprehensive information on CVs, see
\citet{warner95-1} and \citet{hellier01-1}.

Once the white dwarf has accumulated a certain critical amount of material, a 
thermonuclear runaway is triggered on its surface during which 
the previously accreted material (or slightly less or slightly more) is ejected
into the interstellar space. This event, during which the brightness of the
system increases by 8$-$16 magnitudes, is called a (classical) nova 
eruption \citep{bode+evans12-1,chomiuk+21-1}. The
underlying binary system is not destroyed and recommences the accretion process
within a couple of years 
\citep[e.g.][]{retter+98-2,mason+21-1,murphy-glaysher+22-1},
making this a recurrent event. The length of a classical nova cycle is
estimated to $10^{4}-10^{7}$ yr, depending on the mass of the white dwarf and
the long-term accretion rate \citep{townsley+bildsten04-1,hillman+20-1,hillman+20-2}. 

The ejected material forms an expanding nova shell that is typically not
spherically symmetric, but has a prolate or bipolar shape 
\citep{gill+obrien99-1,naito+22-1} that can also present certain asymmetries
\citep{pavana+20-1} and equatorial or tropical rings 
\citep{slavin+95-1,porter+98-1}. The details of the ejection
process are not yet fully understood.
Some studies suggest the occurrence of several ejection events with 
different energetics and 
dynamics \citep[e.g.][]{aydi+20-12,steinberg+metzger20-1,shen+quataert22-1},
with shocks between those different layers of material being responsible for 
the clumpy structure of the ejecta \citep{likwanlok+17-1,harvey+20-1}. Other
cases can be explained by a single ejection event and intrinsic clumpiness
\citep[e.g.][]{liimets+12-2,mason+18-1}. The long-term light curves of nova
eruptions show a large diversity \citep{strope+10-1}, which might reflect
differences in the ejection mechanisms in different novae 
\citep[e.g.][]{mason+20-1}.

The nova \object{V1425 Aql} was discovered on February 2, 1995, by 
\citet{nakano+95-2} as an object of eighth magnitude, although it is likely that
the real maximum with a visual magnitude of $m_v = 6-7$ mag was missed by
two to three weeks \citep{masoncg+96-4,kolotilov+96-4,kamath+97-1}. 
In time-series photometric data in the I-band taken over a range of four months
in 1996, when the nova had declined by approximately 7 mag, \citet{retter+98-2}
detected two main periodic signals at 6.14 h and 1.44 h, as well as the 
beat period between them. These authors interpreted the longer period corresponding to
the orbital modulation, and the shorter period as the spin period of the white 
dwarf, making the system an intermediate polar. However, the non-detection
of any modulation in X-ray data prompted \citet{worpel+20-2} to speculate that
this might be a misclassification, although they also note that the source
was too faint to draw any firm conclusions.

Spectroscopic observations of the ejecta in the first months after the 
eruption showed evidence of a clumpy structure and the presence of optically 
thin dust and a hot ionised gas component with a low dust-to-gas ratio 
$\le10^{-3}$
\citep{masoncg+96-4,kamath+97-1}. \citet{lyke+01-2} estimated the ejected
mass to $2.5-4.2 \times 10^{-5} M_\odot$, which is well within the typically observed
range \citep{tarasova19-1}. \citet{ringwald+98-2} reported the detection
of the shell in optical passbands from narrow-band imaging with the
Hubble Space Telescope (HST) approximately three years after the eruption and 
mention 
the possibility of a bipolar structure in the [O{\sc iii}] data, whereas 
\citet{sahman+15-1} did not find any evidence for a shell in the H$\alpha$ data
taken in 2014 as part of the Isaac Newton Telescope H$\alpha$ Survey (IPHAS).

We obtained photometric and spectroscopic data of V1425 Aql as part of a 
project to determine H$\alpha$ and [O{\sc iii}] luminosities of 
old nova shells \citep[see][]{tappert+20-1} and present our analysis of these 
data in the subsequent sections.

\section{Observations and data reduction \label{obs_sec}}

\subsection{Pre-imaging \label{preima_sec}}

\begin{figure}
\includegraphics[width=\hsize]{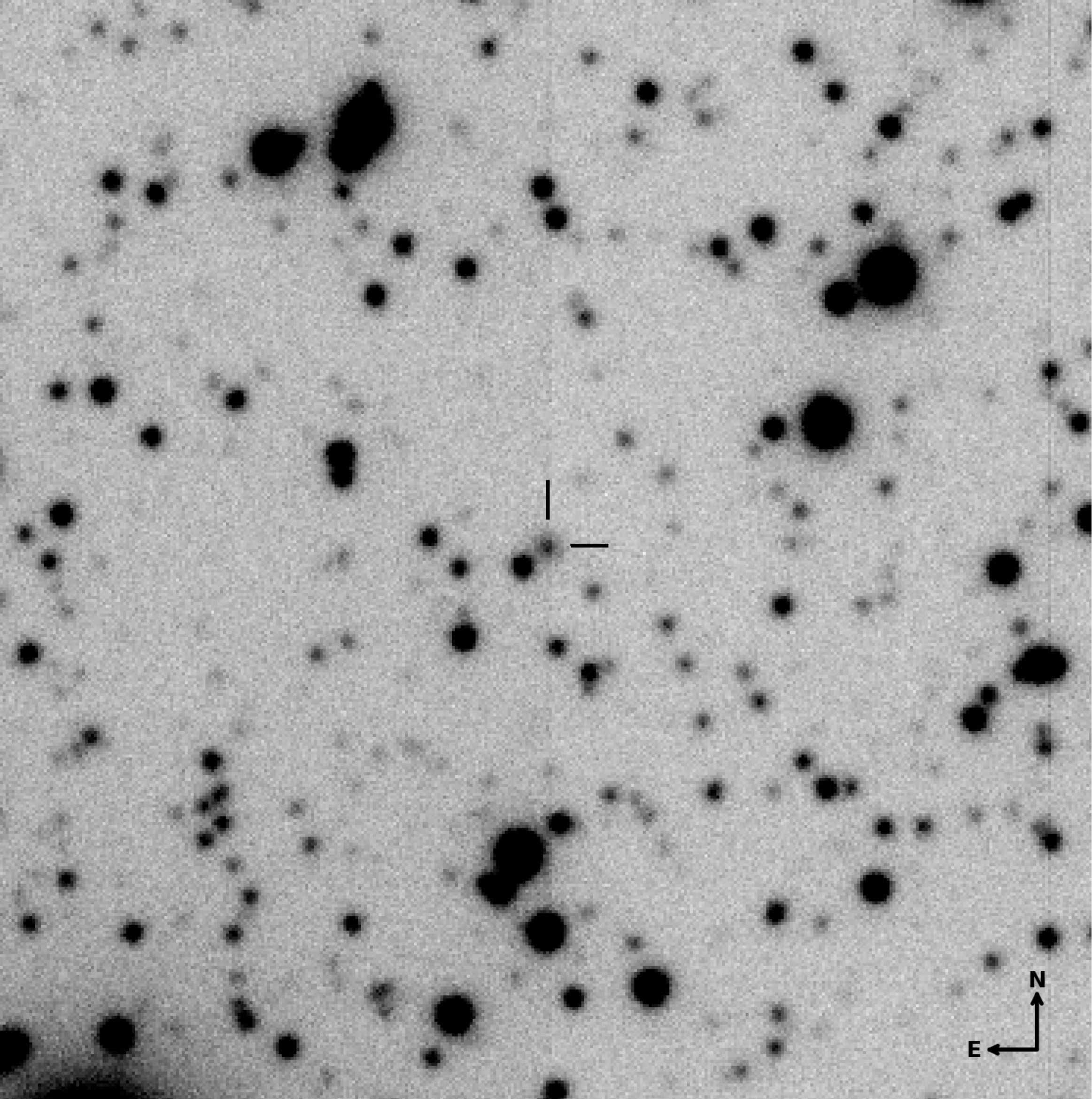}
\caption{Finding chart for V1425 Aql, based on the r$^\prime$ image and centred
on the position of the nova. The size is 1$\times$1 arcmin and the orientation
is such that east is to the left and north is up, as indicated in the lower
right corner.}
\label{fc_fig}
\end{figure}

Because V1425 Aql cannot be unambiguously identified on the finding 
chart from
the \citet{downes+05-1} catalogue, images were taken on July 12, 2018, in the 
r$^\prime$ (G0326) and the H$\alpha$ (G0336) passbands installed on the
Gemini Multi-Object Spectrograph \citep[GMOS;][]{hook+04-1,gimeno+16-2}
at the Gemini-South telescope on Cerro Pach\'on, Chile, in preparation for the 
subsequent spectroscopy. One image was taken with each filter, with
exposure times of 10 s for r$^\prime$ and 90 s for H$\alpha$. Bias subtraction
and flat-field reduction were performed using the Gemini reduction package
installed in the Image Reduction and Analysis Facility (IRAF) software
\citep{tody86-1,tody93-1}. A number of comparison stars were measured in each
frame to correct for a small positional shift between the two images and the
intensity of the H$\alpha$ image was scaled to roughly correspond to the
value of the r$^\prime$ data. Finally, subtracting the r$^\prime$ image from
the H$\alpha$ one unambiguously revealed the nova as the brightest H$\alpha$
source in the field. In Fig.~\ref{fc_fig}, we present a finding chart with
the identification of the nova.

We used the Graphical Astronomy and Image Analysis Tool (GAIA\footnote{%
\url{http://star-www.dur.ac.uk/~pdraper/gaia/gaia.html}}) in combination with
the catalogue of sources identified in the Gaia mission
\citep{gaia16-1} to obtain an astrometric solution for the r$^\prime$ image.
As mentioned in \citet{tappert+20-1}, the nova itself is not a Gaia source.
Its position was measured as the centre of the point spread function (PSF) to
\begin{equation}
\alpha_\mathrm{J2000} = 19^\mathrm{h}05^\mathrm{m}26^\mathrm{s}.631(09),
~~\delta_\mathrm{J2000} = -01^\circ42'03".26(13),
\label{pos_eq}
\end{equation}
which presents an offset of approximately 11 arcsec with respect to the position 
recorded in \citet{downes+05-1}, mainly due to differences in
declination.

\subsection{Spectroscopy\label{spec_sec}}

\begin{figure*}
\includegraphics[width=\hsize]{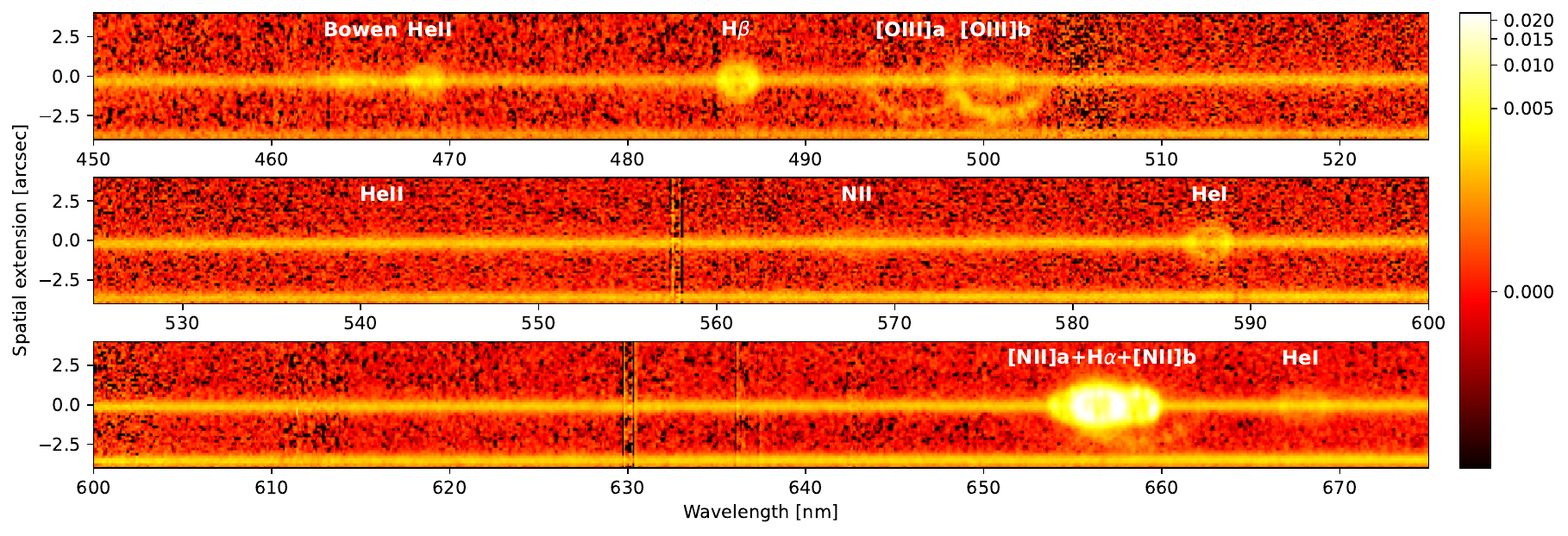}
\caption{Two-dimensional spectrum of V1425 Aql. The identified emission lines
have been labelled. The spatial orientation is such that the y-axis goes
from southwest to northeast at an angle of $223^\circ.3$ north to east.
The zero point of the spatial axis has been set to the position of the
stellar continuum at a wavelength of $\lambda =$ 580 nm. The
colour map is in units of $10^{-18}$ W m$^{-2}$. Its scale is logarithmic in order
to show the details of faint and bright features simultaneously.}
\label{fullsp_fig}
\end{figure*}

\begin{figure}
\includegraphics[width=\hsize]{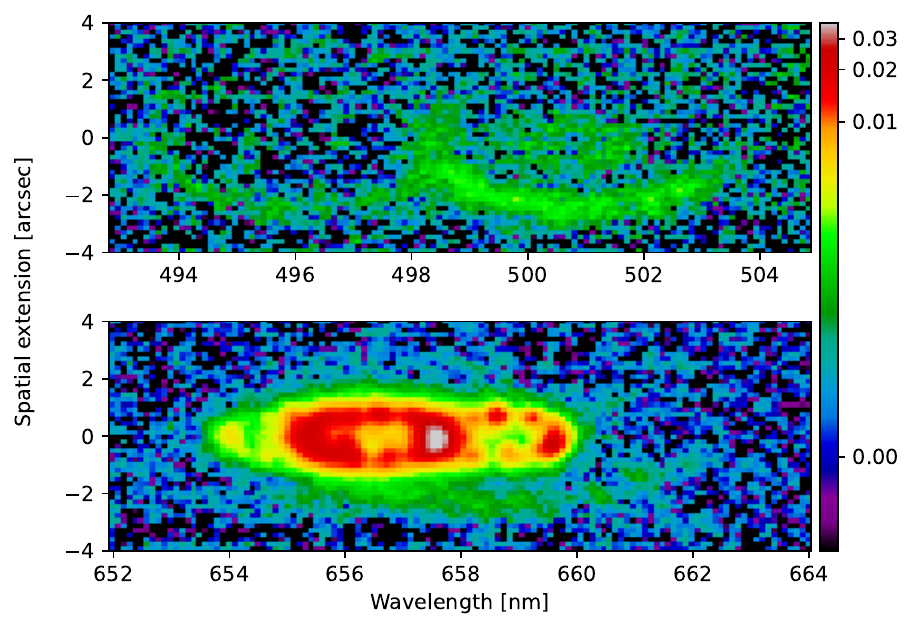}
\caption{Spectral ranges around [O{\sc iii}] (top) and H$\alpha$ (bottom)
with the stellar continuum subtracted. We refer to Fig.~\ref{fullsp_fig} for details of the spatial axis.}
\label{nocont_fig}
\end{figure}

\begin{figure*}
\includegraphics[width=\hsize]{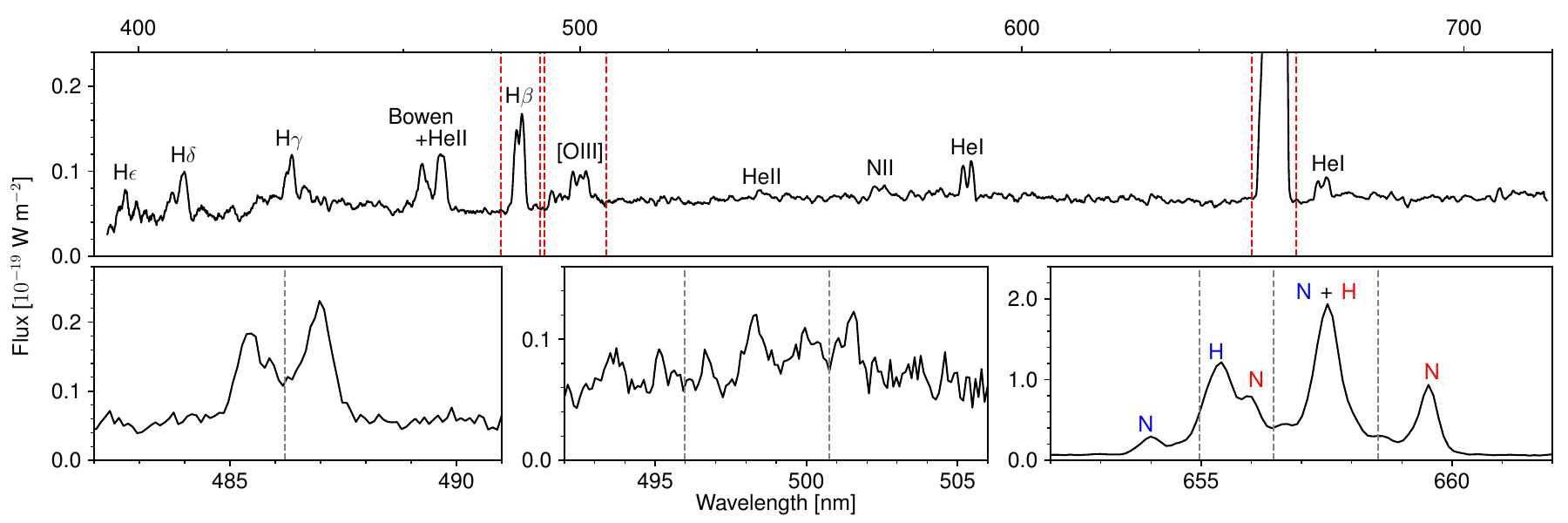}
\caption{Extracted spectrum. The upper plot presents the full available
wavelength range. The data were smoothed with a filter of  7 pixels in width.
Identified emission lines are labelled. 
The lower plots show close-ups of
the unsmoothed data in the regions indicated in the upper plot by the two 
vertical dashed lines (H$\beta$, the two [O{\sc iii}] lines, and H$\alpha,$ plus
the two [N{\sc ii}] lines, from left to right). The vertical dashed lines 
indicate the nominal centres (rest wavelength plus systemic velocity) of the 
respective emission lines. In the rightmost plot, the letters mark the blueshifted 
and redshifted [N{\sc ii}]a,b and H$\alpha$ emission line components of the 
inner shell, and the colour of the letters symbolises the direction of the 
shift.
}
\label{1dspec_fig}
\end{figure*}

Long-slit spectroscopy was performed on September 5, 2018, using GMOS and the
B600+G5323 grating. A 1.0 arcsec slit was centred on the previously
determined position (Eq.~\ref{pos_eq}) with a position angle of
$223^\circ.3$ (measured anticlockwise from north) to avoid contamination 
from the object located at approximately 1.7 arcsec southeast of the nova 
(Fig.~\ref{fc_fig}). Two series of three spectra were taken at two
different central wavelengths (550 and 560 nm) to account for the
gaps in the spectral coverage caused by GMOS using a mosaic of three
CCDs. 
The individual exposure times were 400 s, resulting in a total integration
time of 40 min.
A 2$\times$2 binning was employed, yielding
a spectral resolution of 0.45 nm, measured as the full width at half maximum
(FWHM) of the spectral calibration lines, and a spatial step size of
0.16 arcsec per pixel. From the acquisition image taken in the r$^\prime$
filter, the FWHM of the PSF was measured as $\sim$0.5 arcsec.

As with the other data, the Gemini IRAF routines were applied for bias
subtraction, flat-field correction, wavelength calibration, and correction
for the quantum efficiency differences between the three CCDs. Subsequently,
low-order functions were fitted to the sky background and subtracted.
Flux calibration was obtained by comparison with the spectrophotometric 
standard LTT 7987 that was observed on the same night using a 5.0 arcsec slit. 
The resulting individual images were then averaged to yield a single spectrum
(Fig.~\ref{fullsp_fig}). The respective areas around specific emission lines 
were extracted so that low-order polynomials could be fitted to the stellar 
continuum and subsequently subtracted, which allows us to study the nova shell 
in detail. Examples are presented in Fig.~\ref{nocont_fig}. Finally, a one-dimensional spectrum at the position of the binary was 
established using the IRAF optimal extraction routine \citep{horne86-1};
it is shown in Fig.\,\ref{1dspec_fig}.

\subsection{Narrow-band imaging\label{nb_sec}}

\begin{table}
\caption{Parameters of the narrow-band imaging from July 25, 2019. For each
filter, we give the wavelength range $\Delta\lambda$ in nm
(obtained from the GMOS website), the number of frames $n$, the individual 
exposure time $t_\mathrm{exp}$ in s, and the resulting total exposure time 
$t_\mathrm{tot}$ in min.}
\label{nbanddet_tab}
\centering
\begin{tabular}{lllll}
\hline
\hline\noalign{\smallskip}
Filter & $\Delta\lambda$ & $n$ & $t_\mathrm{exp}$ & $t_\mathrm{tot}$ \\ 
\hline\noalign{\smallskip}
[O{\sc iii}]  & 496.5--501.5 & 23 & 155 & 59.42 \\\hspace{0cm}%
[O{\sc iii}]C & 509.0--519.0 & 5  & 65  & 5.42 \\
H$\alpha$     & 654.0--661.0 & 29 & 35  & 16.92 \\
H$\alpha$C    & 659.0--665.0 & 10 & 31  & 5.17 \\
\hline
\end{tabular}
\end{table}

\begin{figure*}
\includegraphics[width=\hsize]{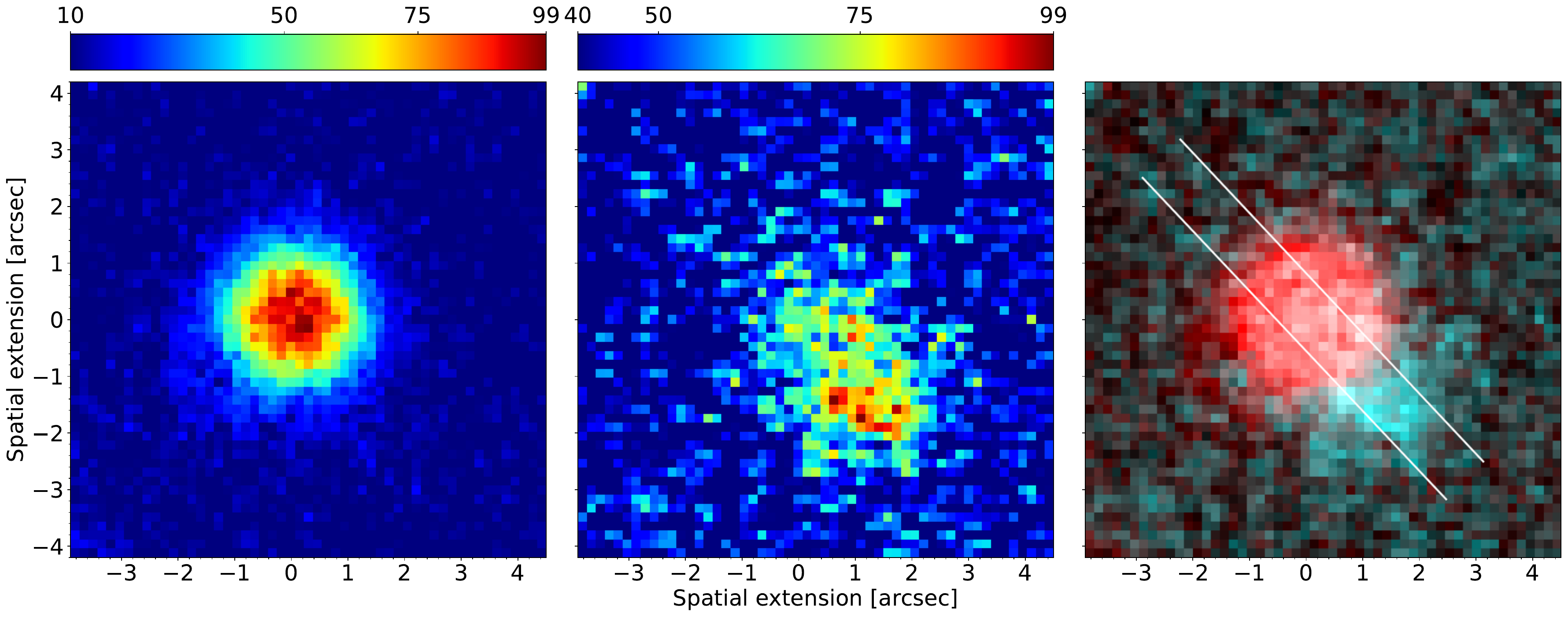}
\caption{Narrow-band images in the usual orientation with east to the left
and north to the top. The left and middle plots show the H$\alpha$
and the [O{\sc iii}] data, respectively. In both images, the sky background
has been subtracted and the data have been normalised 
to a value of 100
with respect to their
individual maximum values. The right plot shows a composite image, where the H$\alpha$ data are marked in red and the [O{\sc iii}] data in blue. 
The white lines indicate the width and angle of the slit in the 
spectroscopic data. For all images, the zero point of the coordinate system 
was set to the centre of the H$\alpha$ emission.}
\label{nband_fig}
\end{figure*}

On July 25, 2019, a series of narrow-band images were taken with GMOS using 
H$\alpha$ (G0336) and [O{\sc iii}] (G0338) filters as well
as filters that cover the corresponding nearby continuum wavelength ranges
(G0337 and G0339, hereafter H$\alpha$C and [O{\sc iii}]C, respectively).
Exposure times were kept comparatively short to avoid saturation. Further
details are given in Table \ref{nbanddet_tab}. 

The Gemini IRAF routines were used for basic reduction, incorporating
bias subtraction and flat-fielding. After correction of the positional
shifts, all images corresponding to a specific filter
were combined into a single frame using a $\pm 1\sigma$ rejection limit.
The instrumental fluxes of a number of comparison stars were measured 
to determine the scaling factors to be applied to the continuum data so that 
they
match the fluxes in the emission line filters. The such scaled continuum frames
were subtracted from the emission line data. The resulting emission 
distributions are given in Fig.~\ref{nband_fig}.
Measuring the FWHM of the PSF for several stars on the images, we derive a
value of $\sim$1.1 arcsec for all filters.

Although the narrow-band images share an identical relative coordinate system,
it proved necessary to perform an astrometric correction in order to obtain
absolute positions. This was done 
in the same way as for the r$^\prime$-Band image (Section \ref{preima_sec}).
The root mean squares of the coordinate fits ranged from 0.04 to 0.09 arcsec,
depending on the S/N.

\section{Analysis\label{ana_sec}}

\subsection{General spectral appearance}

A rough first look at the spectrum (Fig.~\ref{fullsp_fig}) shows an 
unusual shell 
configuration, with a high-velocity component that is confined to the 
southwestern part of the sky. This latter is detected only in the two [O{\sc iii}] emission lines at 
$\lambda$495.9 nm and $\lambda$500.7 nm,  and is somewhat weaker
in [N{\sc ii}] $\lambda$658.4 nm. Additionally, there is a symmetric low-velocity 
component visible both in forbidden and allowed transitions. For 
the sake of brevity, we mark the blue and red lines of the forbidden 
transitions with `a' and `b', respectively, as defined in Table \ref{linepars_tab}.
For similar reasons, throughout the paper we refer to the symmetric
low-velocity ejecta as the `inner shell' and to the asymmetric high-velocity
ejecta as the `outer ejecta'.

Overall, V1425 Aql presents a very rich spectrum (Fig.~\ref{1dspec_fig}). Apart
from the aforementioned forbidden lines, it includes the Balmer hydrogen series 
from H$\alpha$ down to H$\epsilon$, He{\sc i} $\lambda\lambda$667.8, and 567.8 
nm, He{\sc ii} $\lambda\lambda$541.2, and 468.6 nm, and the Bowen blend. The
feature between He{\sc ii} $\lambda$541.2 and He{\sc i} $\lambda$567.8 is
most likely N{\sc ii} $\lambda$568.0 nm. At its red side, there is potentially
a very weak remnant of the once strong [N{\sc ii}] $\lambda$575.5 nm emission
\citep{kamath+97-1,lyke+01-2}, but its intensity is too close to the noise level
to identify it with any certainty. A comparison of Figs.~\ref{1dspec_fig} and 
\ref{fullsp_fig} shows that most, if not all, emission lines are present in the 
ejecta. The only doubtful case is
that of the Bowen blend, which, despite being almost as bright overall as the neighbouring He{\sc ii} line, shows a significantly smaller spatial
extension, if any. At the very least, this line is clearly dominated by
emission from the binary, either from the accretion disc or as a recombination
from the illuminated side of the secondary star. For the other emission lines,
the contribution from sources inside the binary appears to be mostly negligible.
In the He{\sc i} lines, the blueshifted and redshifted components are
sufficiently separated that it should be possible to distinguish any emission
from the binary, but no such feature can be detected with any confidence. In
the bluer lines, such as H$\beta$ and He{\sc ii}, this is more difficult, but
it is clear that the emission from the ejecta is dominant. Many
post-novae appear to have high mass-transfer rates 
\citep[e.g.][]{fuentes-morales+21-1}, implying the presence of a bright, 
optically thick accretion disc, where most emission lines are weak,
and V1425 Aql appears to fall into that category. In such systems, the 
strongest disc emission
line is usually found to be H$\alpha$. Unfortunately, the strong blend of the
H$\alpha$ + [N{\sc ii}] emission from the ejecta (Fig.~\ref{1dspec_fig}, right) 
prevents detection of such a component.
Finally, we note that the spectrum does not present any obvious absorption
lines.

\subsection{Velocities and spatial extensions\label{vel_sec}}

\begin{table*}
\caption{Parameters of the ejecta emission lines identified in the spectrum of
V1425 Aql. For each transition, we give the rest wavelength $\lambda_0$,
the systemic velocity $v_s$, the maximum radial velocity $v_r$, the projected
spatial extension $r_p$, the extinction corrected
flux in the slit $F_s$, and the extrapolated total flux $F_t$. 
The parentheses give the estimated 1$\sigma$ uncertainties.}
\label{linepars_tab}
\centering
\begin{tabular}{lllllll}
\hline
\hline\noalign{\smallskip}
Line & $\lambda_0$ & $v_s$ & $v_r$ & $r_p$ & $F_s$ & $F_t$ \\
 & [nm] & [km s$^{-1}$] & [km s$^{-1}$] & [arcsec] & [$10^{-18}$ W m$^{-2}$] & [$10^{-18}$ W m$^{-2}$] \\
\hline\noalign{\smallskip}
He{\sc ii}    & 468.57 & 44(15)   & 455(21)  & 0.81(08) & 3.69(07)  & 8.21(16)\\
H$\beta$      & 486.13 & 46(15)   & 523(21)  & 0.82(08) & 7.36(13)  & 16.36(28)\\\hspace{0cm}%
[O{\sc iii}]a & 495.89 & 172(14)  & 1508(20) & 2.32(11) & 2.83(11)  & 8.08(30)\\\hspace{0cm}%
[O{\sc iii}]b & 500.68 & & & & & \\
inner         &        & 51(14)   & 448(20)  & 0.72(08) & 2.04(08)  & 4.54(18)\\
outer         &        & 85(14)   & 1497(20) & 2.27(11) & 7.78(14)  & 22.22(39)\\
He{\sc ii}    & 541.2  & --       & --       & --       & 0.17(01)  & 0.37(02)\\
N{\sc ii}     & 568.0  & --       & --       & --       & 0.84(02)  & 1.86(05)\\
He{\sc i}     & 587.57 & 61(12)   & 535(17)  & 0.84(08) & 1.96(04)  & 4.36(08)\\\hspace{0cm}%
[N{\sc ii}]a  & 654.81 & 103(11)  & 479(16)  & 0.66(09) & 3.60(14)  & 8.00(31)\\
H$\alpha$     & 656.28 & 111(11)  & 565(16)  & 0.79(08) & 25(1)    & 55(2)\\\hspace{0cm}%
[N{\sc ii}]b  & 658.36 & & & & & \\
inner         &        & 91(11)   & 483(16)  & 0.68(08) & 10.28(40) & 22.8(1)\\
outer         &        & [75(25)] & 1661(27) & 2.42(11) & 2.41(05)  & 6.88(13)\\
He{\sc i}     & 667.81 & 81(11)   & 559(16)  & 0.85(08) & 0.77(07)  & 1.72(15)\\
\hline
\end{tabular}
\end{table*}

The first attempt to determine velocities by fitting the emission line 
profiles with
Gaussian functions proved unsuccessful because of several overlaps
and blends both in the spatial and the spectral direction,
especially in the H$\alpha$/[N{\sc ii}] complex (Fig.~\ref{1dspec_fig}, right).
For example, within our spectral resolution, the blueshifted component of 
[N{\sc ii}]b overlaps exactly with the redshifted one of H$\alpha$. 
Overall, the blending is such that there
is no transition for which both the blueshifted and the redshifted components could 
be clearly distinguished. In addition, at least for H$\alpha$, a contribution
from the accretion disc or other locations in the underlying binary can be
expected. The bluer emission lines are mostly
not as affected by blending as the region around H$\alpha$, but they present
significantly lower S/Ns (Fig.~\ref{1dspec_fig}, left). In most lines,
this results in too many degeneracies to fit both emission line components,
which would be necessary to calculate first the systemic and then the expansion 
velocities. 

We therefore determined the wavelengths of
the emission lines by drawing ellipses to represent the two-dimensional 
emission distribution of a given line and registering the points that coincide 
with the stellar continuum. The corresponding 3$\sigma$ uncertainties are 
estimated to be in
the order of $\pm$1 pixel, which is equal to $\pm$0.1 nm. 
The systemic velocity $v_s$ was determined as the centre of the two
Doppler-shifted velocity components and with respect to the rest wavelength of 
that 
line. Finally, the radial velocity was 
calculated as the difference between the maximum observed velocity and $v_s$. 
The values for the measured lines are collected in Table \ref{linepars_tab}.
For the outer [N{\sc ii}]b ejecta, the blue Doppler shift could not be 
determined with any sufficient precision. Therefore, in this case, we calculated the
radial velocity of the red component with respect to the mean systemic velocity.
The latter is obtained as the average of the individual values, excluding the
outlier from the [O{\sc iii}]a line, to $v_s = 75(25)$ km/s. In addition, we
derive the average radial velocities for the shell of the allowed
transitions, $v_\mathrm{r,a} = 527(44)$ km/s, for the inner shell of the 
forbidden lines, $v_\mathrm{r,f,i} = 470(19)$ km/s, and for the outer ejecta,
$v_\mathrm{r,f,o} = 1555(92)$ km/s. This yields the ratios
\begin{equation}
\frac{v_\mathrm{r,f,o}}{v_\mathrm{r,a}} = 2.95(30) ~~
\mathrm{and}~~
\frac{v_\mathrm{r,f,o}}{v_\mathrm{r,f,i}} = 3.31(24),
\label{velrat_eq}
\end{equation}
which overlap within one sigma.

We tested the validity of our method by fitting the profiles of the He{\sc i}
$\lambda$587.6 nm emission line components with a Gaussian, because these
are the most isolated and least distorted lines. We obtained a systemic
velocity of 59 km/s, and find the velocities of the lines to be 513 km/s, which lies
well within the uncertainty range of the values determined by the comparison
with ellipses above.

Next, we measured the extensions $r_p$ of the ejecta in the spatial direction.
We calculated the sum of the central three to five CCD columns of a given 
shell line.
The centres of the brightness distributions of the shell and
of the stellar continuum were
measured by fitting Gaussian functions to them and the extension of the shell
was calculated with respect to the latter value. With the precision we achieve here, we find the
inner shell to be symmetrical around the stellar continuum, and so the extensions
into both directions were averaged to yield the values in Table \ref{linepars_tab}.
From the typical FWHM of the Gaussians, we estimate a 1$\sigma$ uncertainty of
two-thirds of a pixel. We find average spatial extensions
of $r_\mathrm{p,a} = 0.82(02)$ arcsec for the shell of the allowed transitions,
$r_\mathrm{p,f,i} = 0.69(03)$ arcsec for the inner shell of the forbidden lines,
and $r_\mathrm{p,f,o} = 2.33(08)$ arcsec for the outer ejecta. The ratios are then
\begin{equation}
\frac{r_\mathrm{p,f,o}}{r_\mathrm{p,a}} = 2.84(12) ~~
\mathrm{and}~~
\frac{r_\mathrm{p,f,o}}{r_\mathrm{p,f,i}} = 3.40(19),
\label{spatrat_eq}
\end{equation}
which overlap within two sigma. From this comparison, there is therefore a possibility
that the inner shell of the forbidden lines and the inner shell of the allowed lines
occupy slightly different velocity and spatial regimes. Given that the two 
transitions need different density conditions, this is not entirely unexpected.

Comparing Eqs.~\ref{velrat_eq} and \ref{spatrat_eq}, the velocity
ratios agree very well with their respective extension ratios. 
Dividing the velocity by the radius ratios, we find a value of 1.04(11) for the
comparison of the outer forbidden lines with the allowed lines and 0.97(09)
for the comparison of the outer and inner forbidden lines. For a nova age
of 23.6 yr, we can still assume a constant expansion velocity without
deceleration \citep[e.g.][]{santamaria+20-1} and therefore the above ratios are also
valid for the time that has passed since the ejection of the material. As
both values agree well with unity within their errors, we can conclude that
both the outer ejecta and the inner shell(s) originate in the same nova 
eruption, with the respective uncertainties translating to time ranges of
2.6 and 2.1 yr for the outer forbidden to allowed and for the
outer to the inner forbidden line regions, respectively.

\subsection{Brightness distribution\label{bright_sec}}

The narrow-band photometry corroborates our findings from spectroscopy.
While the H$\alpha$ emission is placed symmetrically at the position of the
nova, the [O{\sc iii}] distribution shows an elongated shape, with its maximum
positioned to the southwest of H$\alpha$.
The FWHM of the PSF of the inner shell measures 2.1 arcsec, which when
compared to the value of the stellar PSF (1.1 arcsec, Sect.\,\ref{nb_sec}) 
shows that the emission distribution is resolved.
 
We determined the position of the brightness distribution by measuring the
geometrical centre of the H$\alpha$ image first in a contour plot and 
afterwards by applying the centring routine implemented in the GAIA viewer. 
The position of the [O{\sc iii}] emission is more ambiguously defined because
of its asymmetric 
shape and the considerable noise in the data, which affects even the region of 
the maximum intensity (S/N $\sim$3). We opted for the centre of an
ellipse with a major axis of 1.58 and a minor axis of 0.55 arcsec, which encompasses the southern `head' of the distribution (around $x,y$ = 1,$-$1.5 
in the coordinate system of Fig.~\ref{nband_fig}), where the highest 
intensities are located. 

This yields the coordinates
\begin{equation}
\alpha_\mathrm{J2000} = 19^\mathrm{h}05^\mathrm{m}26^\mathrm{s}.626(01),
~~\delta_\mathrm{J2000} = -01^\circ42'03".41(09)
\label{Hapos_eq}
\end{equation}
for H$\alpha$ and
\begin{equation}
\alpha_\mathrm{J2000} = 19^\mathrm{h}05^\mathrm{m}26^\mathrm{s}.553(06),
~~\delta_\mathrm{J2000} = -01^\circ42'04".98(08)
\label{Opos_eq}
\end{equation}
for [O{\sc iii}]. The two centres are therefore separated by 1.91(10) arcsec, 
which is comparable to the values measured in the spectroscopy 
(Table \ref{linepars_tab}; we point out that the head is not fully included in
the slit; see Fig.~\ref{nband_fig}, right), and form an angle of 
215(7)$^\circ$ east of north.

\subsection{Expansion parallax\label{expar_sec}}

Assuming equal velocities in the line of sight and perpendicular to the
observer, the distance $d$ to the nova can be determined from the radial
velocity $v_r$, the projected expansion of the shell in arcsec $r$, and the
time $\Delta t$ that has passed since the nova eruption, as
\begin{equation}
d = \frac{v_r \Delta t}{\tan{r}}
\label{dist_eq}
.\end{equation}

With $v_r$ and $r$ for the different ejecta as given in section \ref{vel_sec}
and $\Delta t = 23.6$ yr for our spectroscopic data (taking January 21, 1995, 
as an 
estimated date for the nova eruption), we obtain an average of $d =$ 3.2(3) 
kpc from the allowed lines, 3.4(3) kpc from the forbidden lines of the inner 
shell, and 3.3(3) kpc from the outer ejecta, yielding a final value of 
$d = 3.3(3)$ kpc. 

Previous distance estimations were provided by \citet{masoncg+96-4} and
\citet{kamath+97-1} as $d = 3.6-4.8$ kpc and $d = 2.7(6)$ kpc, respectively.
Both employed the maximum magnitude rate of decline relation, whose 
usefulness is nevertheless still hotly debated 
\citep{schaefer18-2,selvelli+gilmozzi19-1,dellavalle+izzo20-1}.
The difference in those two distance estimations is mainly due to the different
extinctions used by the authors, with \citet{masoncg+96-4} correcting for
$E_\mathrm{B-V} = 0.55$ mag and \citet{kamath+97-1} using $E_\mathrm{B-V} = 
0.76$ mag. The Stilism
website\footnote{\url{https://stilism.obspm.fr/}} \citep{lallement+19-1}
gives $E_\mathrm{B-V} = 0.67^{+0.04}_{-0.10}$ mag for a limiting distance
of 2.4 kpc in the direction of the nova, which favours the higher extinction
value and therefore the lower distance.

The potential flaw in our distance estimate lies in the
assumption that the measured radial velocity corresponds to the measured
spatial extension, while in reality the two quantities concern two different
parts of the ejected material. As mentioned in Sect. 1, nova
shells are typically not spherically symmetric and a proper distance
determination would require a model for the spatial extension in the line of
sight. We also did not take any possible deceleration of the material into 
account. However, we note that the velocity of the outer ejecta lies well
within the range of velocities observed shortly after the
eruption \citep{masoncg+96-4,kamath+97-1,arkhipova+02-1}, and so this is unlikely
to be a significant effect.

\subsection{Flux\label{flux_sec}}

\begin{figure}
\includegraphics[width=\hsize]{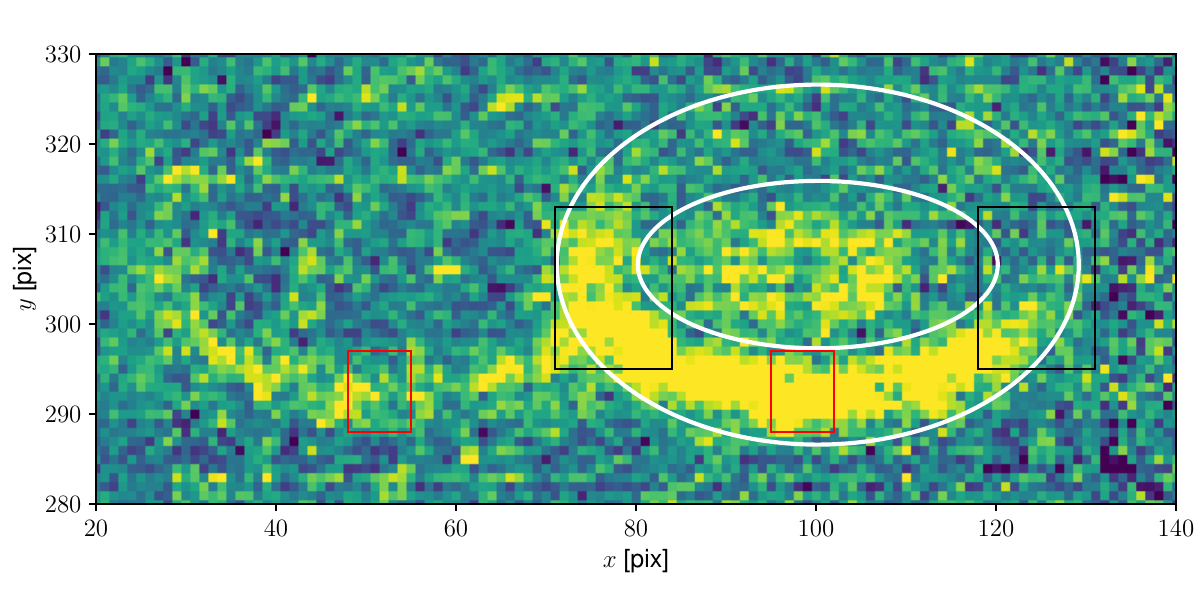}
\caption{Areas used for the flux calculation on the example of the [O{\sc iii}]
lines in the continuum--subtracted spectrum. The ellipses represent the region
used for the elliptical aperture photometry of the outer [O{\sc iii}]b line.
The two red rectangles mark the areas of comparison between the outer 
[O{\sc iii}]a,b lines. The black rectangles indicate the region of overlap of 
the [O{\sc iii}]a line with the outer [O{\sc iii}]b material and the 
corresponding region for [O{\sc iii}]b. For details see the text.}
\label{ellipse_fig}
\end{figure}

For each shell of a given emission line in the 
continuum-subtracted
spectroscopic data, we defined two 
ellipses representing the outer and inner limits of the brightness distribution
(Fig.~\ref{ellipse_fig}).
We then used the EllipticalAperture routine in the Photutils package
\citep{bradley+19-4} within Astropy \citep{astropy13-1,astropy18-1} to
measure the respective total fluxes, calculating the shell flux as the
difference. 
In this way, any potentially present emission from the binary is also
subtracted from the data.
Because the
previous subtraction of the sky leaves the background centred on a zero value,
this also works for the `incomplete` ellipses of the outer ejecta.

Several lines in the spectrum are affected by blending with neighbouring lines,
and therefore the above method is not applicable, at the very least not without 
further
corrections. For the inner ejecta, this concerns the H$\alpha$ + 
[N{\sc ii}] region. Fortunately, it contains two line components that can
be assumed to be mainly free of other contributors: the blueshifted peak of
[N{\sc ii}]a and the redshifted peak of [N{\sc ii}]b (Fig.~\ref{1dspec_fig},
right). 
Measuring the fluxes of the redshifted and blueshifted peak in the H$\beta$ 
emission line, which represents the strongest unblended emission, we find that 
they differ by approximately 5\%. The profile of the line shows that both
peaks are likely composed of several components. In the red peak, these
appear to be closer together in velocity space ---yielding an overall higher 
intensity--- than in the blue peak, where the components are more easily separated
(lower left plot in Fig.~\ref{1dspec_fig}). The other lines of the inner
shell appear to show a corresponding behaviour, including the forbidden lines
as indicated by the albeit more noise-affected [O{\sc iii}] $\lambda$501
emission (lower middle plot in Fig.~\ref{1dspec_fig}). This can be interpreted
as evidence of a certain clumpiness of the inner ejecta. We proceed by assuming
that this structure and the equality of the fluxes in both peaks within
5\%\ are valid for all inner emission lines. 
We now
average the central three rows (roughly corresponding to the stellar PSF) of 
the continuum-subtracted 2D spectrum and fit Gaussian functions
to the four identified peaks, corresponding to the blueshifted component of
[N{\sc ii}]a, the combination of the blueshifted peak of H$\alpha$ and the
redshifted peak of [N{\sc ii}]a, the blend of the redshifted peak of H$\alpha$
with the blueshifted peak of [N{\sc ii}]b, and finally the redshifted peak
of [N{\sc ii}]b, all of which refer to the inner shell. We subtract the
[N{\sc ii}]a,b contribution to the H$\alpha$ line, which is measured on the 
respective
undisturbed components. The results for both H$\alpha$ peaks agree well with
each other and we find an average flux for one H$\alpha$ peak of 
$0.209(10) \times 10^{-18}$ W/m$^{2}$. The ratios to the fluxes of the 
[N{\sc ii}] lines are H$\alpha$/[N{\sc ii}]a = 6.95 and H$\alpha$/[N{\sc ii}]b 
= 2.40. 

We repeat the procedure for the inner shell of [O{\sc iii}]b, measuring 
only the better-defined red peak. This yields flux ratios 
[N{\sc ii}]b/[O{\sc iii}]b = 11.11 and [N{\sc ii}]a/[O{\sc iii}]b = 3.84. 
Because the flux for [O{\sc iii}]b can be measured using the elliptical
aperture photometry, we can now use these factors to calculate the total 
fluxes contained in the slit for the two [N{\sc ii}] lines from the 
[O{\sc iii}]b flux. The flux of the H$\alpha$ line then follows from the 
H$\alpha$/[N{\sc ii}] ratios.

The discernible emission lines of the outer ejecta, [O{\sc iii}]a, 
[O{\sc iii}]b, and [N{\sc ii}]b, also present different degrees of overlap. The
most severely affected line is that of [N{\sc ii}]b, the blueshifted half
of which
is hidden within the H$\alpha$ and [N{\sc ii}] emission from the inner shell.
Thus, we calculate the fluxes by comparing the intensities of the three lines
of the outer ejecta using the elliptical aperture photometry of the outer 
[O{\sc iii}]b line as the base flux, which we first have to correct for the 
overlap with the [O{\sc iii}]a line on its blue side.
We assume that all three emissions track the same material and that
the intensity ratios are constant over the entire spatial extension. Then,
for all three lines, we can define rectangular regions that correspond to the
same respective velocity and spatial range, and where all three lines are 
undisturbed from other contributors (for the [O{\sc iii}] lines; these are the 
red rectangles in Fig.~\ref{ellipse_fig}). Similar to the elliptical aperture
photometry, the size of those areas is not important as long as they contain 
the same emission portions for all three lines and exclude other 
emission sources, because the background is centred around zero.
Comparison of the total intensities contained in these areas yields the ratios 
[O{\sc iii}]b/[O{\sc iii}]a = 2.86(10) and [O{\sc iii}]b/[N{\sc ii}]b = 
1.47(01). The uncertainties were estimated by displacing the areas by 1 pixel
in each direction and repeating the measurement. 

To obtain the total flux in the slit for the [O{\sc iii}]b line, we can use
its elliptical aperture photometry, but we have to correct for the overlap with
[O{\sc iii}]a. We define two rectangular areas (black rectangles in 
Fig.~\ref{ellipse_fig}), the first one including the complete overlap region, 
that is the red part of [O{\sc iii}]a and the blue part of [O{\sc iii}]b. The 
second area defines the red part of [O{\sc iii}]b, which corresponds to the same
velocity and spatial range. From the intensities contained in these areas and
using the above intensity ratios, we can now calculate the contribution from
[O{\sc iii}]a to the blue part of [O{\sc iii}]b and correct the flux from the 
elliptical aperture photometry by subtracting it. The total slit fluxes of 
[O{\sc iii}]a and [N{\sc ii}]b are then calculated from that value divided
by their respective ratios.

We estimate the total emitted flux by first rotating the narrow-band
images with respect to the position angle of the spectroscopy and sum
the instrumental flux of the area that corresponds to a 1.0 arcsec slit. A
comparison with the total instrumental flux shows that in the case of 
H$\alpha$ the slit contains a fraction of 0.45 of the total flux, and in the 
case of [O{\sc iii}] it contains a fraction of 0.35. In the second step, we sum the rows 
of the spectroscopic data that contain emission from the shell, use the 
SpectRes Python package \citep{carnall17-1} to rebin the spectra and the 
normalised filter curves of the four narrow-band filters (see Section 
\ref{nb_sec}) to a common step size, and finally multiply the spectrum with 
each filter curve to calculate the included flux. This yields fluxes of 
$F_\mathrm{H\alpha} = 8.45$, $F_\mathrm{H{\alpha}C} = 2.85$, 
$F_\mathrm{[OIII]} = 0.89$, and 
$F_\mathrm{[OIII]C} = 0.76 \times 10^{-18}$ W/m$^2$ for the four filters, and 
therefore the differences $F_\mathrm{H\alpha}-F_\mathrm{H{\alpha}C} = 5.60$ and 
$F_\mathrm{[OIII]}-F_\mathrm{[OIII]C} = 0.13 \times 10^{-18}$ W/m$^2$, which can now 
can be compared to the instrumental fluxes of the narrow-band imaging. Assuming
that the above factors of 0.45 and 0.35 determined for H$\alpha$ and for 
[O{\sc iii}] can be used in general for the inner and the outer 
ejecta, respectively, we calculate the total fluxes $F_t$ for each emission 
line and shell.

As the final step, we use the Astropy dust$\_$extinction 
package\footnote{\url{https://dust-extinction.readthedocs.io/en/stable/index.html}} 
to correct the fluxes for the interstellar extinction according to
\citet{fitzpatrick04-1}. We apply
$E_\mathrm{B-V} = 0.76$ mag, because this appears to be the best
estimate at present (see Section \ref{expar_sec}). The resulting
fluxes are shown in Table (\ref{linepars_tab}).

Comparing the fluxes of the forbidden lines, we note that the two ejecta
present almost opposite [N{\sc ii}]/[O{\sc iii}] ratios, because 
$F_\mathrm{t,[NII]b} / F_\mathrm{t,[OIII]b} =$ 0.31(01) for the outer material, 
but 5.08(28) for the inner shell.

\subsection{HST data\label{hst_sec}}

\begin{figure}
\includegraphics[width=\hsize]{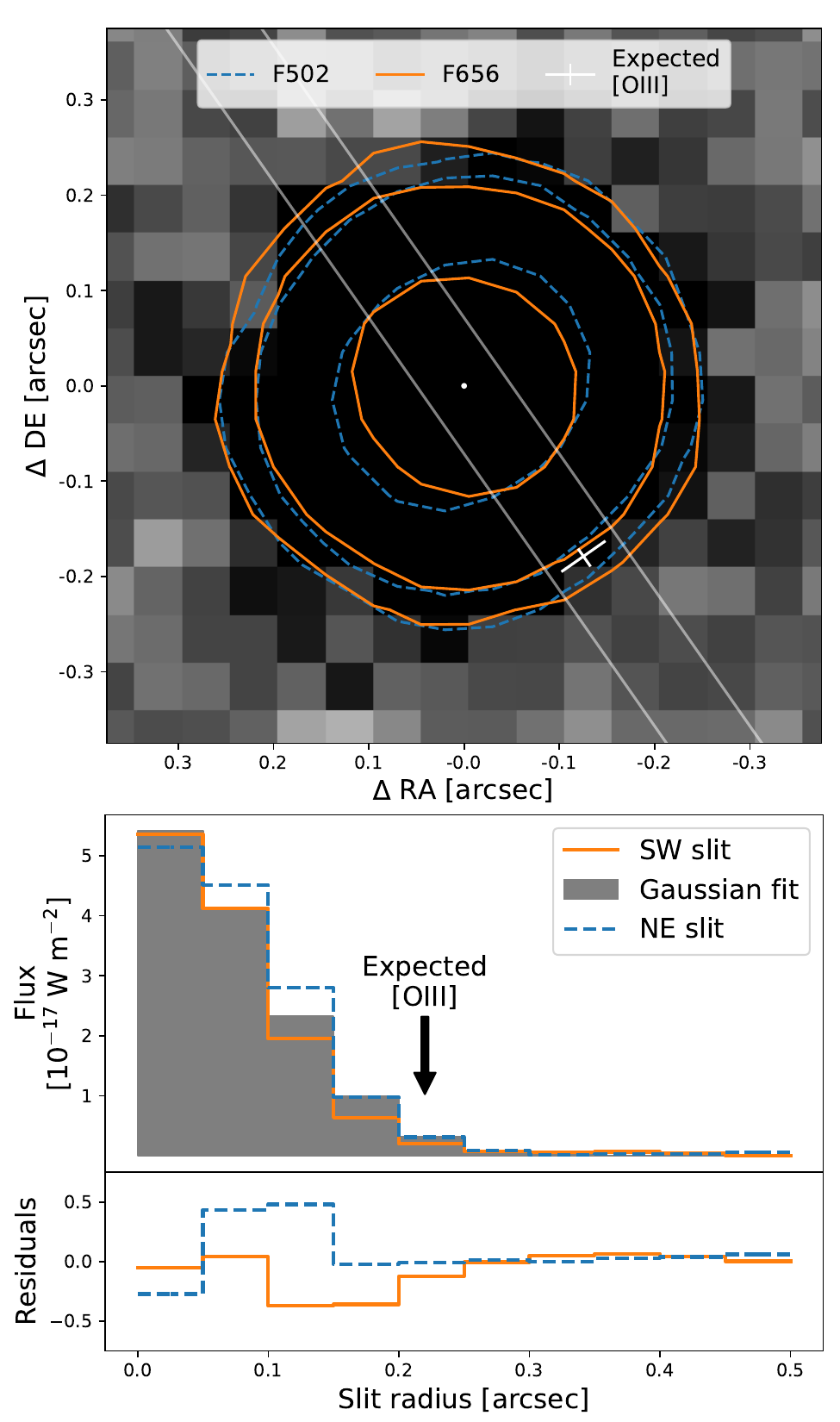}
\caption{Flux distribution of the HST data. The top plot shows the contours of 
the F656 and the F502 flux overlaid on the F656 image, representing fractions 
of 0.5, 0.1, and 0.05 of the respective central 
fluxes. The F502 data have been corrected for the small instrumental shift (see 
text). The white cross marks the expected position of the outer [O{\sc iii}] 
ejecta, with the sizes of the marker denoting the uncertainty in the radial 
distance and angle. The lines mark the area used to derive the
radial profile shown in the middle plot. Dashed and solid lines
represent the flux in the two directions towards and away from the expected
outer ejecta, while the filled area indicates a Gaussian fit. The bottom plot
presents the respective residuals to that fit.
}
\label{hst_fig}
\end{figure}

Hubble Space Telescope data for V1425 Aql were collected using the Wide 
Field Planetary Camera instrument (WFPC2) on November 19, 1997, almost three 
years after the nova event. A total of four images were taken in filters F469, 
F502, F656, and F658. \citet{ringwald+98-2} present a brief summary of the data, 
suggesting that the F502 image, which includes the [O{\sc iii}] line, could 
show a bipolar structure with a 0.2 arcsec separation. 

We reanalysed these archival data to determine whether it is possible to 
observe the 
asymmetric ejecta already at such an early stage. The target was located in 
the field of view of the planetary camera of the instrument, which has a pixel 
scale of 0.0455 arcsec/pixel. The image had been calibrated to have physical 
units of W m$^{-2}$ nm$^{-1}$ using the standard procedure for 
WFPC2 images. We used the width and pivot wavelength of the 
filter\footnote{\href{http://svo2.cab.inta-csic.es/theory/fps/index.php?id=HST/WFPC2-PC.F502N&&mode=browse&gname=HST&gname2=WFPC2-PC\#filter}{HST F502 Filter Service}} 
to obtain physical units of W m$^{-2}$ arcsec$^{-2}$ and to 
determine AB magnitudes. Aperture photometry with a radius of 0.5 arcsec and 
annuli of 0.5 and 1.5 arcsec for the inner and outer radii for background 
subtraction yields an AB magnitude of 12.61(1) mag for V1425 Aql in the F502 
filter. 
The FWHM of the stellar PSF is measured as $\sim$0.15 arcsec, while that of
the nova amounts to $\sim$0.21. Thus, similar to the GMOS data, the
nova shell is just, but clearly, resolved.

At the time of the GMOS NB observations ($\Delta t = 24.5$ yr), the 
asymmetric ejecta are located at a distance of 1.91(10) arcsec from the centre 
of the H$\alpha$ emission (Section \ref{bright_sec}). Assuming a constant 
expansion rate, we obtain a value of 0.078(4) arcsec/yr. 
Thus, at 
the time of HST observations, the asymmetric ejecta should be at 0.22(01) 
arcsec from the position of the nova.

Figure~\ref{hst_fig} shows the F502 image with the overimposed contour levels of 
the F656 filter. Originally, the two different emission distributions presented
a small but noticeable offset of 0.025 arcsec between their
centres. Measuring the positions of one of the stars in the two images, we 
find a very similar offset. We can therefore attribute this shift between filters 
to an instrumental effect and have corrected this correspondingly for the plot.
There, the position where we expect to observe the asymmetric ejecta is marked 
with a cross.

We note that the two emission distributions occupy the same area,
and that they can be well described by a 
circle, especially at lower intensities. The only exception is in the centre of the F502 emission, which presents
a certain elongation. However, this latter lies roughly perpendicular to the direction in which
the outer ejecta would be expected and is therefore unlikely to be connected with it.

We test for a possible excess in flux at the expected position of the ejecta 
by comparing the radial profile of the F502 distribution in the direction of the
ejecta and in the opposite direction. For this purpose, we define a `slit' area 
in 
which we sum up the pixel values as a function of the distance from the centre
of the distribution (middle and bottom plot in Fig.~\ref{hst_fig}). We conclude
that no excess emission is observed near that position.

To investigate the evolution of the inner shell, we compare the HST F656 data
with the GMOS H$\alpha$ image. For this, we first have to find common ground
for the respective astrometries, and must therefore refer to the
r$^\prime$-band image (Fig.~\ref{fc_fig}), because GMOS H$\alpha$ and F656 do
not have any common stars. We find one star in the F656 image that is (barely)
not saturated in r$^\prime$ and can be used for comparison. This star is also
in the Gaia Data Release 3 catalogue, which lists its coordinates as 
$\alpha_\mathrm{J2000} = 19^\mathrm{h}05^\mathrm{m}23^\mathrm{s}.833,
~~\delta_\mathrm{J2000} = -01^\circ42'53".07$ and proper motions as
$-0.871(20)$ and $-2.422(18)$ mas/yr in right ascension and declination,
respectively. Correcting for the proper motion that corresponds to the time
difference between F656 and r$^\prime$ observations, we find offsets
of $\Delta \alpha_\mathrm{F656-r^\prime} = 0.51$ arcsec and 
$\Delta \delta_\mathrm{F656-r^\prime} = 0.08$ arcsec. The difference between
the GMOS r$^\prime$-band and H$\alpha$ images is negligible, because both
astrometries were corrected with respect to the Gaia catalogue. Taking the above
shift into account, we now find that the centre of the GMOS H$\alpha$ emission
differs from that in the F656 observation by $\Delta \alpha_\mathrm{F656-H\alpha}= 
0.12$ arcsec and $\Delta \delta_\mathrm{F656-H\alpha} = 0.29$ arcsec, amounting
to 0.31 arcsec in total, and in the direction of 201$^\circ$ east of north.

Finally, we employ aperture photometry to determine the fluxes $F$ in the two 
filters. Using a large aperture of 0.65 arcsec (compare to Fig.~\ref{hst_fig})
yields $F(\mathrm{H\alpha}) = 1.53 \times 10^{-16}~\mathrm{W/m}^2$ and 
$F(\mathrm{[OIII]}) = 1.17 \times 10^{-16}~\mathrm{W/m}^2$, and, when corrected
for interstellar extinction, $F(\mathrm{H\alpha}) = 7.92 \times 
10^{-16}~\mathrm{W/m}^2$ and  $F(\mathrm{[OIII]}) = 1.33 \times 
10^{-15}~\mathrm{W/m}^2$. In the following section, we show how these fluxes
relate to previously determined values and the GMOS data.

\subsection{Flux evolution\label{fluxevol_sec}}

\begin{table}
\caption{Flux data. With the exception of the final two data rows, which
were derived in this study, all values were taken from 
\citet{downes+01-2}.\label{flux_tab}}
\centering
\begin{tabular}{lll}
\hline
\hline\noalign{\smallskip}
$\log(\Delta t)$ & $\log(F(\mathrm{H\alpha}))$ & $\log(F(\mathrm{[OIII]}))$ \\
\relax
[yr] & [W m$^{-2}$] & [W m$^{-2}$] \\
\hline\noalign{\smallskip}
$-$0.854 & $-$12.25 & $-$13.51 \\
$-$0.854 &          & $-$13.41 \\
$-$0.699 & $-$12.37 & $-$13.17 \\
$-$0.509 & $-$13.11 & $-$13.40 \\
   0.364 & $-$15.74 & $-$15.49 \\
   0.450 & $-$15.76 & $-$15.57 \\
   1.373 & $-$16.97(02) & $-$18.40(02) \\
\hline
\end{tabular}
\end{table}

\begin{figure}
\includegraphics[width=\hsize]{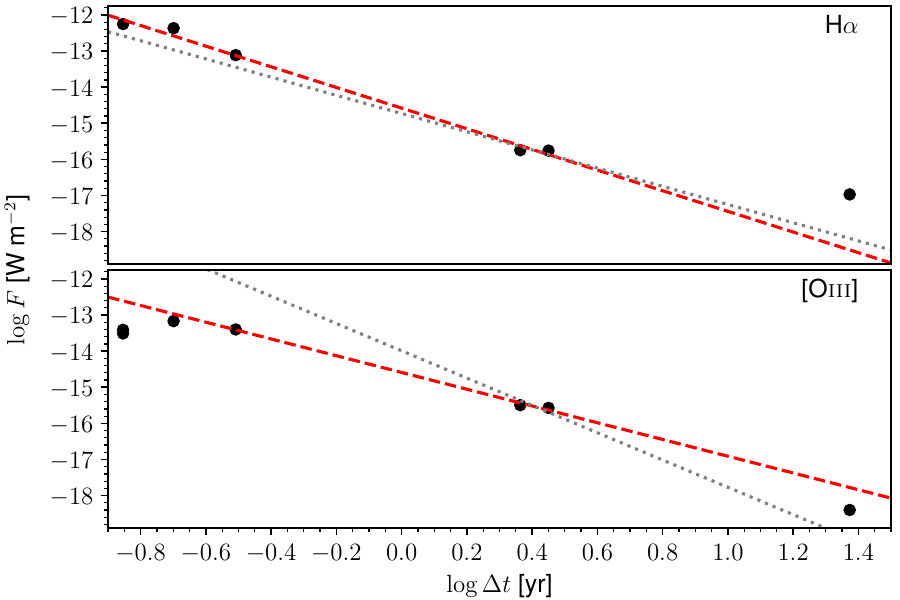}
\caption{H$\alpha$ (top) and [O{\sc iii}] flux vs.\,time on a double logarithmic
scale. The dashed lines represent fits to the three data points within $-0.6 
< \log(\Delta t) < 0.6$, while the dotted lines were fitted to the two points
near $\log(\Delta t) = 0.4$ with fixed respective slopes of the S-class from 
\citet{tappert+20-1}.
}
\label{fluxevol_fig}
\end{figure}

The evolution of the luminosity of nova shells was originally investigated
by \citet{downes+01-2} and later refined by \citet{tappert+20-1} who
limited the sample to systems with a well-known distance and grouped the
data according to speed class and to the light-curve classification
scheme introduced by \citet{strope+10-1}. 
The authors defined three different phases in the long-term evolution of
the nova shell flux for both the H$\alpha$ and the [O{\sc iii}] emission:
a roughly constant behaviour in the initial and 
late stages, and a linear decline slope  in between (on a double logarithmic 
scale). These three stages were interpreted as corresponding to the transition
from optically thick to optically thin material, the free expansion phase,
and the start of the interaction with the interstellar medium (ISM), 
respectively. The slopes and 
time intervals that define the three sections appear to vary somewhat 
with respect to light curve type and/or speed
class, but most 
data sets
suffered from incomplete sampling, leaving room for
significant systematic uncertainties.

Because V1425 Aql has not been detected by Gaia, it was not part of the
\citet{tappert+20-1} study, although \citet{downes+01-2} incorporated
the flux data from \citet{kamath+97-1} in their catalogue. With the above 
fluxes derived from the HST and GMOS observations, the data on V1425 Aql now 
cover
a much larger time range. For a comparison of the decline slopes, we do not
need the conversion to luminosity, because this only results in a different 
vertical zero point. Therefore, we can avoid the uncertainty
related to distance and interstellar extinction. Because the flux values in
Table \ref{linepars_tab} are already extinction corrected, we collect all
the `raw' flux data used in this section in Table \ref{flux_tab}. As
we show in Sect.\,\ref{inexdisc_sec}, we can assume that emission from the 
outer ejecta
was below the detection limit in the HST data. For consistency, we limit
the analysis of the flux evolution to the inner shell.

The data are plotted in Fig.\,\ref{fluxevol_fig}. According to 
\citet{strope+10-1}, V1425 Aql is a member of the S-class, which presents smooth
decline light curves. For this class, \citet{tappert+20-1} find that for
H$\alpha$, the initial, roughly constant, stage corresponds to a time interval
of $\log(\Delta t) = [-2.0,-1.0]$, the decline stage to $[-1.3,1.4],$ and the
late, again roughly constant, stage corresponds to $[0.9,1.8]$. For 
[O{\sc iii}], the
corresponding values are $[-1.2,-0.4]$, $[-0.5,1.2]$, and $[0.9,1.7]$, respectively.
The fact that some of these ranges overlap is evidence for the aforementioned 
incomplete sampling. Unfortunately, the available data on V1425 Aql suffer
from the same problem, presenting two large gaps from approximately 0.4 to 2 yr 
and from 3 to 20 yr. Taking the above ranges for the S-class, for [O{\sc iii}], 
the data before the first gap should correspond to the initial `constant' 
phase, the data in between the gaps to the decline, and the final data point 
(the GMOS flux) to the late `constant' phase. For H$\alpha$, all data before 
the second gap should be part of the decline, whereas the last data point 
possibly corresponds to the late stage. However, as mentioned above, those 
$\Delta t$ limits are not very well defined. We therefore decided to fit two 
functions to each data set. The first is fit to three data points: 
the last one before the first gap and the two before the second gap. 
This yields slopes of $-2.86(22)$ for H$\alpha$ and of $-2.32(12)$ for 
[O{\sc iii}]. The second function is a fit to the latter two data points with 
the slope fixed to the value for the S-class, namely $-2.52(06)$ and 
$-3.78(28)$, respectively. 

The patchy sampling of both the \citet{tappert+20-1} and the V1425 Aql 
data certainly necessitates a large amount of caution in interpreting the results.
Still, we find that they at least fit into a consistent picture. For 
H$\alpha$, the two slopes overlap within 1.5$\sigma$, and also the first two 
data points not included in the fit agree well with the resulting
slope. However, the GMOS flux lies clearly above that decline. All
data points before the second gap can therefore be regarded as being part of the decline
phase, while the GMOS observations already show the late phase in the H$\alpha$
evolution. On the other hand, the slopes for the [O{\sc iii}] data differ by
almost 5$\sigma$. It is therefore likely that all data before the first
gap still belong to the constant phase and so the inclusion of the last one 
of those points in the fit yields a slope that is too shallow. Assuming a steeper
slope, as for H$\alpha$, the GMOS [O{\sc iii}] flux could also already 
correspond to the late phase, although the offset from the decline slope is 
not as pronounced. Fitting only the three data points after the first gap 
yields a slope of $-2.95(16)$, which overlaps with the S-class slope within 
3$\sigma$.

\section{Discussion\label{disc_sec}}

\subsection{One or two shells?\label{onetwodisc_sec}}

At first glance, the V1425 Aql data suggest the presence of two distinctive 
ejecta that cover different velocity and spatial regimes. We find that they 
originate in the same nova eruption, but our precision leaves a time range of 
approximately $\pm$2 years between the actual ejection events. During 
recent years,
evidence has accumulated that ---at least in some novae--- the ejection of 
material in the eruption is not a single event, but takes place in stages and 
involves different ejection velocities \citep{friedjung11-1,chomiuk+21-1}.
The typical picture is that of an early slow ejection in the equatorial plane
and a subsequent high-velocity bipolar outflow that can occur up to several
hundred days later \citep{woudt+09-2,chomiuk+14-1}.
We emphasise that our time resolution, based on the velocities and distances,
allows for both a single ejection scenario and for two ejections with
a time difference of up to approximately two years. We also note that the 
long-term
light curve of the nova provided by the American Association of Variable Star
Observers (AAVSO)\footnote{\url{aavso.org}} within this time range presents a 
scatter (at times up to slightly more than 1 mag) and a coverage (including 
two gaps with a length of $\ge$ two months) that would allow a minor or
even a significant rebrightening ---that could have accompanied a second ejection---
to remain undetected. Furthermore, the light curve in \citet{strope+10-1},
although classified as `smooth' between $\Delta t \sim$60 and 90 d, presents a 
clear bump with a size of approximately 0.5 mag. Such bumps and other 
fluctuations 
occur in many novae, including some systems of the S-class, and are not conclusive evidence for additional ejections. However, their 
largely
unknown origin leaves room for a number of possibilities \citep[e.g.][who 
explain jitters and oscillations in nova light curves by unstable nuclear 
burning on the WD]{mason+20-1}.

The two emission regions clearly present different densities, with that of the
outer one being too low to enable allowed transitions. In principle, this
distribution could also represent a single asymmetric shell. Indeed, we note
that the offset of 0.31 arcsec of the centre of the H$\alpha$ emission with 
respect to the HST data is at an angle similar to the location of the outer 
[O{\sc iii}] material (Section \ref{hst_sec}), which could reflect a general 
tendency of the ejecta to move in that direction. However, the FWHM of the 
GMOS H$\alpha$ shell amounts to approximately 2.2 arcsec and, considering that 
the
astrometric comparison between HST and GMOS is based on a single star, it is
not certain that the observed offset is really significant. Photometry under
good seeing conditions and in a spectral range that avoids the main emission
from the shell could allow us to pinpoint the location of the underlying binary and thus
clarify whether or not the inner shell is symmetrically placed around the
post-nova.

A finding in favour of the interpretation of two ejecta is the fact that both 
the [O{\sc iii}] and [N{\sc ii}] emission distributions present 
a local intensity minimum between the inner and outer emission regions,
both in space and in velocity. This indicates that the emission distribution 
cannot be explained by a gradually decreasing density. The spatial separation
is not discernible in the [O{\sc iii}] narrow-band image, which does not show
any clear presence of the inner shell. However, this is likely due to the 
large flux difference between the two ejecta for this transition (Table 
\ref{linepars_tab}). In fact, the 2D spectrum (Fig.~\ref{fullsp_fig}), 
where the inner [O{\sc iii}] shell is clearly detected, does not reflect the 
real flux difference, because the slit covers a larger fraction of the inner 
shell than that of the outer ejecta. As can be seen in Fig.~\ref{nband_fig}, 
even the maximum of the [O{\sc iii}] distribution is significantly affected by 
noise. Therefore, the apparent absence of the inner shell in that image can be 
explained by low S/N
and the significantly poorer seeing conditions in the narrow-band images compared
with those associated with the spectroscopic data (1.1 arcsec vs.\,0.5 arcsec; Sects.\,\ref{nb_sec}
and \ref{spec_sec}, respectively).

\subsection{Intrinsic or extrinsic?\label{inexdisc_sec}}

\begin{figure}
\includegraphics[width=\hsize]{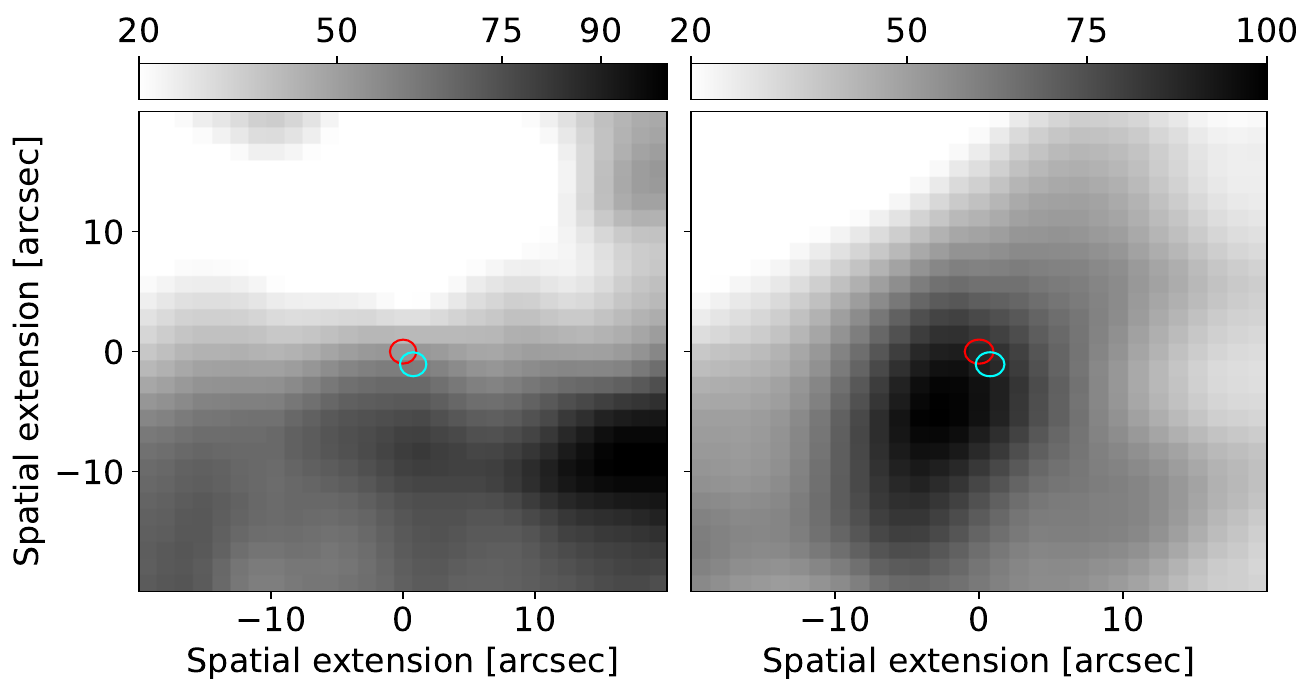}
\caption{WISE data of the 40$\times$40 arcsec field centred on V1425 Aql.
The left and right plots show the band 3 and band 4 images, respectively. As in the other 
maps, the orientation is north up, east to the left. Both linear colour maps 
have an intensity range of 
20\%\ (white) to 100\%\ (black) of the intensity range in the shown area.
The circles mark the positions of the two ejecta. The size of the circles also 
roughly corresponds to the extension of the ejecta.}
\label{wise_fig}
\end{figure}

Deviations from axial symmetry in nova shells are well known. For example,
the shell of IPHASX J210204.7+471015 presents a severe asymmetry, including
a strong spatial separation of the H$\alpha$ and the [O{\sc iii}]-emitting 
material \citep{santamaria+19-1}, and \citet{schaefer+19-1} list a number of 
other examples. However, these concern mainly comparatively old novae
(IPHASX J210204.7+471015 has a suspected age of 130 to 170 yr),
where the observed asymmetries are likely caused by interaction of the
ejecta with inhomogeneously distributed ISM.
On the other hand, it has been speculated that intrinsic non-axisymmetric
ejections could be common and that they could play an important role in the evolution of
CVs, causing temporary orbital eccentricities with consequences for the
mass-transfer process \citep{nelemans+16-1,schaefer+19-1}.

Therefore, in principle, there are two possibilities: either the observed asymmetry 
of the outer ejecta is due to interaction with asymmetrically distributed ISM,
or it is due to an intrinsic effect, meaning an asymmetric ejection mechanism.
To investigate the former possibility, we examined data from the
Wide-field Infrared Survey Explorer \citep[WISE; ][]{wright+10-2}, which is available 
in four wavelength bands centred on 3.4, 4.6, 12.1, and 22 $\mu$m, 
respectively. Of special interest for our case are the latter two, with
band 3 being sensitive to hot ISM and band 4 sensitive to cold dust. 

The band 3 data indeed present an asymmetric distribution of ISM
close to the position of the nova, with an intensity gradient that rises
from north to south (left plot of Fig.~\ref{wise_fig}). However, because we do
not know the distance of that feature, it is unclear as to whether it is really
in the vicinity of the nova or whether this is just a projection effect.
Still, let us assume for a moment that this ISM band indeed causes the 
originally (axis-)symmetrically ejected material to be seen as a comparatively
isolated blob. As the gradient increases towards the outer 
ejecta, the asymmetry would therefore have to be caused not by obscuration of
one part of the ejecta, but by excitation; 
that is, in the northeastern part, the ISM is not sufficiently dense to
interact with the ejecta, and thus the latter remains invisible, while in
the southwestern direction, the ejecta plough into the more
concentrated ISM, with part of the shock energy being transformed into
radiation.
However, this interpretation has several problems. First, while
the gradient is increasing towards the south, there is still material present
north of the nova, and so it is hard to imagine that there would be no interaction
at all in that region. Second, even if the extension of the inner shell is
only approximately one-third of that of the outer ejecta, some effect of 
asymmetric ISM should still be visible, which is not the case. 
Finally, this possibility requires the radiation from the outer ejecta to
be exclusively caused by interaction with the ISM.
From the flux evolution (Sect.~\ref{fluxevol_sec}), we find indeed that such 
interaction is likely to contribute to the flux for the inner shell, and thus
even more probably to the outer ejecta.
However, the
distance of the asymmetric blob to the binary amounts to approximately 
$1\times10^{12}$
km $=$ 0.032 pc, using $v_\mathrm{exp} = 1550$ km s$^{-1}$ and $\Delta t =
23.6$ yr. At that distance, the material should still be affected by
photoionisation \citep[compare e.g.\,with the shell of V382 Vel, which has
a radius of $\approx$0.045 pc; ][]{takeda+diaz19-1} and it is not clear why
this should affect only one side of a hypothetically symmetric ejecta.

The WISE band 4 data, which are sensitive to colder material, do not show any enhancement
in the northeastern direction that could obscure emission from ejecta in
that region either. Instead, there is a concentration at a distance of
approximately 3.4 arcsec southeast of the nova; it is unlikely to be related 
to the
nova, because if it were a remnant of a previous nova eruption at a
---again hypothetical--- former position of V1425 Aql, one would expect a roughly
ring-like distribution, which is very different from the observed one. In any 
case, the
location of the centre of the distribution makes it unlikely that
it causes the asymmetry in the outer ejecta.

Looking for an intrinsic origin for the asymmetry, we need to identify a
mechanism that provides an isolated and well-defined location on the surface
of the white dwarf. The one process that immediately springs to mind is
accretion along the magnetic field lines that deposits the material from the 
donor star onto one of the magnetic poles. As mentioned in Sect. 1,
some observations identify V1425 Aql as an intermediate polar 
\citep{retter+98-2}, whereas others suggest that the magnetic field is
rather weak \citep{worpel+20-2}. In the latter case, it is unlikely that
it would have any effect on the nova eruption \citep{livio+88-2}.

Finally, the HST data (Sect.\,\ref{hst_sec}) have the potential to
distinguish between an extrinsic and an intrinsic origin, because they were
taken at a significantly earlier time after the eruption than the GMOS data,
but still at a point where the separation of the two ejecta would have already 
been discernible. However, the fact that the inner shell is detected 
but the 
outer ejecta are not means the case remains inconclusive. Considering 
that the GMOS data show the outer material to be significantly brighter in 
[O{\sc iii}] than the inner shell (Table \ref{linepars_tab}), the ratio of the 
[O{\sc iii}] flux of the two ejecta must have been reversed at some point 
between the two data sets. In principle, a possible reason for this could be 
that the F502 filter data of the inner shell contain a strong contribution
from the continuum of the binary, which would not be present in the outer 
ejecta.  However, the spectra presented by \citet{kamath+97-1} and 
\citet{lyke+01-2}, both taken roughly 2.3 yr after the eruption ---which is
approximately half a year before the HST data---, indicate that the filter will 
almost 
exclusively have sampled the emission of the two [O{\sc iii}] lines. Therefore,
for the flux ratio to become inverted, the [O{\sc iii}] flux of the inner shell must have suffered a steeper 
initial decline than that of the outer ejecta and/or the latter must have 
experienced an increase at a later stage.
As the GMOS flux lies below the detection limit of the HST observations,
we cannot decide which scenario applies. From our analysis of the flux 
evolution (Sect.\,\ref{fluxevol_sec}), the GMOS [O{\sc iii}] emission of the 
inner shell could possibly already include
a contribution from interaction with the ISM. Because the outer shell has
covered a significantly larger distance, one could speculate that it
experiences a stronger effect of this kind.

Also, the assumption of a constant expansion rate of the outer ejecta 
could be wrong.
Instead, it could have been ejected a considerable
amount of time later than the inner shell, as has been observed for the 
(symmetric) high-velocity material in V445 Pup \citep{woudt+09-2} and V959 Mon 
\citep{chomiuk+14-1}, meaning that at the time
of the HST observations, it was still too close to the inner shell to be
resolved. This implies that its velocity  was initially larger than the current
value and consequently experienced a braking due to interaction
with the ISM. 
Still, this would require the presence of sufficiently dense
ISM in that region, for which we find no indication from the WISE data. We also
note that velocity-wise deviations from free expansion are
typically expected to occur for
novae that are much older than V1425 Aql \citep{duerbeck87-4,santamaria+20-1}.

\subsection{Temperature, density, or abundances?}

From Sect.\,\ref{flux_sec}, we note that the [O{\sc iii}] -- [N{\sc ii}]
ratio is very different for the two ejecta, with [O{\sc iii}] being much more
prominent in the outer material. The transitions of both lines are strongly
dependent on temperature and density \citep{osterbrock+ferland06-1}, and so these
are the parameters that are most likely to be responsible for the difference. On the other 
hand, in principle, the two ejection processes could also have involved 
different mixing, meaning that more material from the inner parts of the white 
dwarf was dredged up for one ejection than for the other, resulting in
different abundances in the ejecta.

In their analysis of data taken approximately 820 d after the eruption, 
\citet{lyke+01-2} find two classes of material with two different densities 
and velocities, which they interpret as evidence for the `clumpiness of the 
ejecta'. Their measured line FWHMs are spread between 550 and 1600 km/s and are 
clearly influenced by mixing from the various line-emitting regions. As 
none of the shell components had yet been resolved, light at different 
velocities and from different shell components is mixed in the PSF. 
Nevertheless, the highest ionised lines associated with a density $n_e$ of 
$3.5\times 10^4 \rm cm^{-3}$ and a temperature of 13\,800\,K show an average 
FWHM of approximately $1500 \pm 100$ km/s, while the neutral lines, which are 
associated with densities of $2.1-4.6\times 10^5 \rm cm^{-3}$ for temperatures 
of 10.000--15.000\,K on average show velocities of approximately 
$1080 \pm 55$ km/s.
These velocities are fairly consistent with our measurements, taking half the 
FWHM as the velocity of a radially symmetric expanding shell and the full FWHM 
as the velocity of a plume or shell that is only ejected in one direction. In 
hindsight, we might therefore interpret the denser `clumps' described by \citet{lyke+01-2}
as the beginning what we now see as the dense, inner shell, while the 
low-density material they measured turned into the outer shell ejected only to 
one side.

We can test how the densities of these two components have evolved and
extrapolate them assuming that no further material was 
added to the shell. This assumption is reasonable as the measurements 
of \citet{lyke+01-2} are from day 820, while the white dwarf nuclear burning
turnoff already occurred on day 400. We also assume that the radius of the 
shell at turnoff defines the thickness of the sphere $D = 400\,\mathrm{d}
\times v_\mathrm{ej}$, which  then continues to freely expand in the radial direction 
with a constant velocity $v_\mathrm{ej}$. We use $v_\mathrm{i} = 500\,\rm km/s$
(i.e.\,roughly half the FWHM) for the slow, high-density shell component and 
$v_\mathrm{o} = 1500\,\rm km/s$ for the fast, low-density component. For a 
constant amount of material, we can use $n_e(8624\,\mathrm{d}) = 
n_e(820\,\mathrm{d})\times V(820\,\mathrm{d})/V(8624\,\mathrm{d})$ for the 
volumes $V$ at the respective times and derive the electron densities at the 
time of the GMOS observations for the two shell components as 
$n_\mathrm{i} = 1.37-3.01 \times 10^3\,\rm cm^{-3}$ and $n_\mathrm{o} = 0.23 
\times 10^3\,\rm cm^{-3}$. Both values are sufficiently low to be considered 
at the low-density limit, which allows us to derive the temperature from the 
respective [O{\sc iii}] and [N{\sc ii}] line ratios 
\citep{osterbrock+ferland06-1}.

Unfortunately, with [O{\sc iii}] $\lambda$436.3 nm and [N{\sc ii}] 
$\lambda$575.5 nm, both ratios involve lines that are not detected in our
data. We derive upper flux limits from the $3\sigma$ variation of 
the background in our spectrum. For [O{\sc iii}], we find $F_\mathrm{i} 
(\lambda436.3)<4 \times 10^{-20}\,\rm W\,m^{-2}$ and $F_\mathrm{o}
(\lambda436.3)<5 \times 10^{-20}\, \rm W\,m^{-2}$, 
which yields upper limits 
for the temperatures of $T_\mathrm{i,[OIII]} < 12\,000$ K and $T_\mathrm{o,[OIII]}
< 7500$ K for the inner and outer shell, respectively. For [N{\sc ii}], we 
derive $F_\mathrm{i}(\lambda 575.5)<4\times 10^{-20}\,\rm W\,m^{-2}$ and 
$F_\mathrm{o}(\lambda 575.5)<3\times 10^{-20}\,\rm W\,m^{-2}$, yielding
$T_\mathrm{i,[NII]} < 5000$ K and $T_\mathrm{o,[NII]} < 7800$ K. 
Although these upper limits
are not conclusive for exploring the differences between the two ejecta,
they are reasonable in that they are below the value derived by 
\citet{lyke+01-2} and above the temperatures expected for older nova shells
\citep[e.g.][]{osterbrock+ferland06-1}.

Both extrapolated densities are too low to have any significant effect
on the forbidden emissions, and so these
depend mainly
on their transition
probabilities. Therefore, the difference in the densities in the two ejecta cannot 
explain the observed difference in the line ratios between [O{\sc iii}] and 
[N{\sc ii}]. In fact, because the latter transition is the `most forbidden' one
\citep{wiese+96-1}, a higher density should have a quenching effect on
[N{\sc ii}] rather than on [O{\sc iii}]. However, in our data, [N{\sc ii}] is 
stronger than [O{\sc iii}] in the denser, inner shell and weaker in the 
outer ejecta. Therefore, we only see two possible causes of this difference: 
abundance or temperature. 
Our data are not suitable for measuring the abundances of either of the 
two shell components. We can only say that if the two components were 
ejected at different moments of the nova eruption, a different mix between
dredged-up and accreted material could have been ejected, and so this
explanation remains a possibility. Concerning the temperature, it is likely
---although unproven because we only have upper limits---
that the inner shell has a lower temperature than the outer ejecta. 
Because the $D_2$ level of [O{\sc iii}] needs a higher energy to be populated
than the corresponding level in [N{\sc ii}], we could be in a position where 
the different temperatures of the shell components simply provide the perfect conditions 
in the sense that the $D_2$ level of [O{\sc iii}] is still populated 
normally in the outer shell but not in the inner one; 
therefore, the [O{\sc iii}] lines are quenched and become weaker. 
We would like to stress that while this is a possible scenario, it cannot 
be supported with our data.

\section{Summary and conclusions\label{sum_sec}}

We present spectroscopic and narrow-band imaging data of the nova
V1425 Aql that show the emission distribution of the ejecta approximately 23 
yr after
maximum brightness. We find that the material ejected during the nova eruption
consists of two distinctively different components: one low-velocity component and one high-velocity
component (referred to throughout the paper as the inner shell and outer ejecta, respectively). 
The former is 
likely
situated symmetrically around the nova and is detected in
the hydrogen, He{\sc i}, and He{\sc ii} emission lines, as well as in the
[O{\sc iii}] and [N{\sc ii}] transitions. There is some indication that the
material seen in the forbidden lines and the material seen in the allowed ones 
occupy slightly different velocities and spatial regimes, but this is not
conclusive, because the values still agree within two sigma.
From the profiles of the emission lines, we find evidence of inhomogeneities
in the ejected material, confirming the findings of \citet{kamath+97-1} and 
\citet{lyke+01-2}. 

The outer material has an elongated shape with maximum brightness 
situated 
1.91(10) arcsec at an angle of 215(7)$^\circ$ east of north from the 
nova. 
The ratios of velocity and spatial extensions for the low-
and high-velocity ejecta are such that we can conclude that they
originated in the same event, although not necessarily at the same time. 
The two ejecta have significantly different [N{\sc ii}] -- [O{\sc iii}]
line ratios, likely due to different temperatures and/or abundances.

With the available data, we are unable to distinguish between an external and
an intrinsic origin of the asymmetry of the high-velocity material. 
Intuitively, the latter appears more likely, but the evidence is inconclusive.
Comparison of the H$\alpha$ and [O{\sc iii}] flux evolution since the nova 
eruption with other novae suggests that some of the emission
is caused by interaction with the ISM. However, we did not find any evidence
for an inhomogeneous distribution of the latter that could explain the
observed asymmetry.

The main purpose of this paper is to draw attention to the existence of this
---to our knowledge--- unprecedented phenomenon. We did not attempt to model
the geometry of the nova ejecta 
\citep[e.g.][]{gill+obrien99-1,ribeirov+13-3}, because recently acquired 
integral field unit spectroscopy with the Multi Unit Spectroscopic Explorer 
(MUSE) will directly provide the 3D distribution of the material. A detailed
analysis of these data will be published elsewhere (Celed\'on et al., in prep.).

Last but not least, we point out that the discovery of these unusual ejecta was
achieved because a close field star was located at such an angle that it forced
us to place the slit in a certain direction that coincided with
the distribution of the ejecta. As deep spectroscopic surveys or deep 
narrow-band imaging studies involving an [O{\sc iii}] filter of medium-aged 
nova shells are rare, it appears possible that similar phenomena could be 
present in other novae as well, which would have significant 
implications for our understanding of the ejection mechanism.

\begin{acknowledgements}
We thank the referee, Elena Mason, for her thorough review and stimulating
discussion, which led to many improvements and broadening of the scope of the
paper.\\

The ejecta of V1425 Aql and nova eruptions in general were topics of many 
pleasant discussions with Steve Shore at the University of Pisa during a brief 
but very fruitful stay of one of the authors (CT). Many thanks for a wonderful 
and insightful time.

We would also like to thank Timo Kravtsov for his valuable suggestions in 
interpreting the emission line ratios.\\
  
CT acknowledges financial support from Conicyt-Fondecyt grant No.~1170566.
LC acknowledges economic support from ANID-Subdireccion de capital 
humano/doctorado nacional/2022-21220607.

Based on observations obtained at the international Gemini Observatory, a 
program of NSF’s NOIRLab, which is managed by the Association of Universities 
for Research in Astronomy (AURA) under a cooperative agreement with the 
National Science Foundation on behalf of the Gemini Observatory partnership: 
the National Science Foundation (United States), National Research Council 
(Canada), Agencia Nacional de Investigaci\'{o}n y Desarrollo (Chile), 
Ministerio de Ciencia, Tecnolog\'{i}a e Innovaci\'{o}n (Argentina), 
Minist\'{e}rio da Ci\^{e}ncia, Tecnologia, Inova\c{c}\~{o}es e 
Comunica\c{c}\~{o}es (Brazil), and Korea Astronomy and Space Science Institute 
(Republic of Korea). Observing IDs: GS-2018B-Q-132, GS-2018B-Q-230,
GS-2019B-Q-130. The data were processed using the Gemini IRAF package.

Based on observations made with the NASA/ESA Hubble Space Telescope,
and obtained from the Hubble Legacy Archive, which is a collaboration
between the Space Telescope Science Institute (STScI/NASA), the Space
Telescope European Coordinating Facility (ST-ECF/ESA) and the
Canadian Astronomy Data Centre (CADC/NRC/CSA)

This research has made use of the NASA/IPAC Infrared Science Archive, which is 
funded by the National Aeronautics and Space Administration and operated by the
California Institute of Technology.

\end{acknowledgements}

\bibliographystyle{aa}

\end{document}